# "Context, Content, Process" Approach To Align Information Security Investments With Overall Organizational Strategy

Pankaj Pandey

Gjøvik University College, Teknologivn. 22, 2815 Gjøvik, Norway

*ABSTRACT*

*Today business environment is highly dependent on complex technologies, and information is considered an important asset. Organizations are therefore required to protect their information infrastructure and follow an inclusive risk management approach. One way to achieve this is by aligning the information security investment decisions with respect to organizational strategy. A large number of information security investment models have are in the literature. These models are useful for optimal and cost-effective investments in information security. However, it is extremely challenging for a decision maker to select one or combination of several models to decide on investments in information security controls. We propose a framework to simplify the task of selecting information security investment model(s). The proposed framework follows the "Context, Content, Process" approach, and this approach is useful in evaluation and prioritization of investments in information security controls in alignment with the overall organizational strategy.*

## 1. INTRODUCTION

Organisations in today's technology-driven world are heavily dependent upon Information Technology (IT) for the core business processes. Information Technology is no longer seen as a pure rationalization tool indeed IT now forms the foundation of core business processes [1]. A large number of frameworks exist to link the information technology strategy with the business strategy, and are commonly known as Strategic Alignment and Business-IT alignment (e.g. [2,3,4]). The models of strategic alignment present alignment of information technology strategy with the business strategy and consequently it is important for the organizations to pursue information security in alignment with the business strategy. Various researchers like [5,6,7,8] have highlighted that investments in information security controls cannot be considered independent of corporate business strategy.

To pursue information security in alignment with business strategy, organizations are expected to establish a balance between the likelihood of an attack on the information infrastructure and the expected cost on risk management [9]. Information security professionals often find it difficult to quantify the expenses, estimate the likelihood of risks and the nature of information security. So, for some of the information security managers, information security is a "bottom-less pit that never gets full" [10] and for some others, information security is a "necessary evil that hinders productivity" [11]. In such a scenario of uncertainty, one of the difficult question for an organization is "how much is really enough for information security?" [12]. In an attempt to answer these questions, some researchers have claimed that there is an optimal spending level for information security [13,14]. These researchers have suggested not to spend above or below the optimal investment level. One of the definition for optimal information security investment is as "one that best utilizes budget resources (be it sufficient or insufficient) to yield the best possible information risk mitigation strategies" [15].

Researchers have proposed a large number of information security investment models based upon economic and financial theories. Despite the common objective of these investment models the foundation of these models is different. For example, there are investment models that are based on classical economic theories, game theory, accounting principles, real options, macro-economics and system dynamics. The difference in information security investment models is not restricted to the foundation theories . Indeed, the models differ in their set of assumptions as well. For example, some models are based on assumption of a risk-averse decision maker while some models are best suited for a risk-seeking or risk-neutral decision maker; some models assume that the investment in information security controls is a one-time in- vestment while other models assume an investment strategy split over a period [16]. The applicability of these investment models also change with the change in the assumptions of the model. Given all these parameters that needs to be considered when selecting an investment model, it is clear that the decision maker is facing a challenging task. Thus, there is a need for a framework to simplify the model selection process by comparing and evaluating the models based on a set of generic and detailed parameters. Therefore, to support the decision making process of investments in information security controls which is in alignment with the overall organization strategy, we present a framework for the simplification of the selection process of investment model(s) based on the 'Context, Content, Process' approach. The proposed framework is then applied to the selected investment models to present the strengths and weaknesses of the selected models.

The remainder of the paper is structured as follows: Section 2 presents the research method followed for the formulation of framework and selection of information security investment models; Section 3 presents the related work; Section 4 presents the framework for simplification of information security investment model selection task; and Section 5 covers a comparison and analysis of selected information security investment models based on the proposed framework. A discussion on the subject and contribution of the paper is presented in Section 6. Conclusion and directions for future research are presented in section 7 and section 8, respectively.

## 2. RESEARCH METHOD

This section presents the research method followed for the collection of the published literature on frameworks for simplification of the model selection process for investments in information security and to compare the various models of information security investments. The first sub-section describes the process of collection of related frameworks. The process of selection of in-formation security investment models is described in the second sub-section. The third sub-section describes the iterative process of formulation of the proposed framework.

### 2.1. Collection of Related Frameworks

A two-step process was followed to identify the published frameworks for the comparison and analysis of information security investment models. First, a systematic search was conducted on Google scholar, IEEE Xplore, Springer and ACM database using keywords: 'comparison of information security investment models', 'framework for investment in information security', and 'analysis of information security investment models', to identify the relevant articles. In this process, we found a large number of articles covering the decision-making process with regards to information security investments. A filtering approach was followed in the next step. The filter was applied to identify the papers presenting a framework for comparison and analysis of investment models. In this two-step process, only three relevant papers were found. In the three papers, frameworks to compare the investment models for information security controls were discussed. Further, the three frameworks are different in their approach.

### 2.2. Collection of Information Security Investment Models

The articles presenting an information security investment model were found through Google search, Google Scholar, Springer, IEEE Xplore and ACM databases. The keywords used for the search were: 'information security economics', 'economics of information security', 'investment in information security', 'investment models in information security', and 'information security investments'. A filtering approach was applied to the search results to identify the articles with information security investment models. Articles, in which a survey or an overview of several models was given, were discarded. However, an extension to an existing investment model was considered as a new model. Articles presenting an investment model applicable in a particular application, for instance for an intrusion detection system or for handling of customer data were included. These information security investment models that are specifically designed for a particular application or service were analysed to identify the underlying assumptions of the model to understand the difference from generic investment models. We restricted our- selves to the investment models published between 2000 and 2014. After this, we analysed the identified articles to identify the evaluation parameters that are in line with the overall business strategy.

### 2.3. Comparison and Analysis of Investment Models

A systematic search was conducted to identify the investment approaches typically adopted in the information technology field. We used the keywords: 'information technology investment framework', 'information system investment framework', 'information technology investment evaluation', and 'information system investment evaluation' to find the investment frameworks in information technology. After this process, various information technology investment frameworks were found. After carefully reviewing the literature on investment frameworks in information technology and information security, we finalized the 'Context, Content, Process' (CCP) approach to compare and analyse the information security investment models. The CCP approach is useful for aligning the information security investment decisions with the organizational strategy.

After this, we applied the proposed framework on six information security investment models and compared them to each other. For each model, the respective article was read carefully, and the information related to frame- work parameters was marked and extracted, as is presented in section 4.

## 3. RELATED WORK

An overview of existing frameworks for comparison and analysis of investment models in information security is presented in this section. The section begins with a discussion on existing frameworks for investment models in information security and ends with an introduction to the 'Context, Content, Process' approach for investments in information systems.

In our systematic literature search to identify the frameworks for comparison and analysis of investment models in information security, only three articles [16,17,18] were found.

The framework presented in [16] is based on the 'context, content, approach' for the evaluation of information security investment models. The framework presents a limited set of criteria to compare the information security investment models. They applied the framework on the three investment models to demonstrate its usefulness in comparison and analysis of information security investment models.

The framework presented in [17] to compare the information security investment models is based on the parameters derived from the policy and security configuration management tool. The parameters which are proposed in their framework to compare the information security investment models are: (i) "Products bought as a whole" i.e. the decision to invest is a one-time decision without an option to defer, (ii) "Financial measures", i.e. the aim of the tool is to reduce the cost for policy and security configuration management, (iii) "Non-financial measures", i.e. the model is also evaluated on non-financial parameters such as trustworthiness, etc., (iv) "One-time costs and benefits", i.e. the costs (of implementation) are considered as costs incurred at a particular time, (v) "Running costs and benefits", i.e. the costs of maintenance and monitoring and benefits such as reduction in cost to identify misconfiguration incurring over a period are taken into consideration, (vi) "Without explicitly considering attacks", i.e. approaches should be applicable without considering information on attacks and (vii) "Network effects", i.e. the benefits of investments shared across the network are also considered.

One of the major limitations of the framework in [17] is that it considers only those information security investment models which are developed for one-time investments. Secondly, the framework does not considers the attack information while comparing the information security investment models despite the fact that investment models differ in their assumption for attack type, i.e. on the basis of single attack type or multiple attack scenario. Thirdly, the framework does not consider the budgetary limitations; practically the budget for information security investments is limited. Therefore, it is important to compare the information security investment models on the basis of budget available for investments in information security controls.

The framework presented in [18] to compare the information security investment models, is based on and is an extension of framework presented in [19]. The framework in [19] is to quantify the uncertainty in the policy based economic models. The framework in [18] used an accounting framework in conjunction with the framework in [19]. This accounting framework was originally developed for formulation and evaluation of policies related to greenhouse emissions [20].

The framework presented in [18] is applied to three different information security investment models to present a comparison and analysis of these models. These investment models were evaluated on the following criteria: (i) Type or form, i.e. the class of investment model and its mathematical structure, (ii) History and previous applications, i.e. the original purpose for which the model was proposed and where it has been used, (iii) Underlying assumption, i.e. the simplifications made to enable easier application, (iv) Decision that the model supports, i.e. the types of decisions which a decision maker can evaluate using the model, (v) Inputs and Outputs, i.e. the quantities and/or qualities on which the model is based, (vi) Parameters and Variables, i.e. elements affecting the output of the investment model, (vii) Applicable domain and range, i.e. the temporal and physical ranges of inputs, outputs, variables and parameters considered by the model, and (viii) Supporting data, i.e. an evidence that the phenomena of interests are presented accurately by the model.

One of the limitations of the framework in [18] is that it uses extremely broad parameters for comparison and analysis of information security investment models. For example, the parameter of 'underlying assumptions' does not explicitly specify the exact factors on which the investment models will be evaluated; a range of assumptions can be made for an information security investment model, such as assumptions related to attack type, costs, etc. Thus, a decision maker interested in using this framework will have to carefully analyse several information security investment models. The process of analysis of various information security investment models for a non-technical decision maker will be a tough task and thus the decision maker will find it difficult to align the information security investments with the overall organizational strategy.

The framework presented by us is able to address the limitation of [18] framework, by providing a list of carefully drawn parameters from the 'Context, Content, Process' perspective. Therefore, a comparison of information security investment models based on the CCP approach is helpful in deciding investments in information security from the overall organizational strategy.

This article addresses the limitations of the frameworks presented in [16,17,18].

## 4. PROPOSED FRAMEWORK

The 'Context, Content, Process' approach was proposed by Pettigrew [21] for organizational change. The 'Context, Content, Process' approach was later adopted in the information systems domain. We reviewed the literature [22,23,24,25,26,27] on application of 'Context, Content, Process' in the information systems domain and the application of CCP in information systems to understand the strategic factors affecting the economic benefits and emphasizes on an integrated evaluation approach. The selection of CCP framework is driven by two reasons. Firstly, CCP is widely accepted in the information systems community. Secondly, the concepts of CCP are broad enough to accommodate the various factors of organizational strategy. Thus, a framework developed on the basis of CCP to evaluate the information security investment models is useful in aligning the information security investment strategy with the organizational strategy.

An evaluation of information security investments from the 'Context, Content, Process' perspective answers the questions of what is being evaluated, by whom and the purpose of the evaluation. The interaction and linking be- tween the 'Context', 'Content' and 'Process' allows an evaluation of information security investments from multiple dimensions. For example, an absence of connection between the Context, Content and Process will ignore the rea- sons of evaluation and the impact of conflict between the stakeholders.

A thorough analysis of the literature on information security investment models and existing frameworks for comparison and analysis of investments in information security and information systems domain helped us in identifying the factors influencing the 'Content', 'Context' and 'Process' in an organizational setup. In the following subsections we present an analysis of the 'Context', 'Content' and 'Process' aspects of an organization's strategy.

### 4.1. Context

The 'Context' aspect focuses on the pressures from within the environment. This includes pressures from both the internal and external environment. The 'Context' aspect identifies the motivation of organizations behind the investments in information security, business drivers, and stakeholders.

We have derived the following parameters from the 'Context' perspective for evaluation of models of investments in information security:

**(1) Cost of Conflict of Interest:** A decision to invest in information security controls involves various stakeholders such as the management and technical team. Therefore, in case of a conflict between the stakeholders, there will be a cost associated with conflict of interest between the stakeholders. Hann and Weber [28] have modelled the conflict of interest between the senior management and the CIO. The cost of conflict of interest between a principal (for instance, a senior manager) and an agent (for instance, the CIO) is known in economics as an agency cost.

29

Agency costs arise in a variety of other situations where the decision making authority is delegated by a principal (for instance, an owner) to an agent (for instance, a senior manager).

**(2) Type of Decision Maker:** The decision making power of decision makers are derived from the organizational policy and business strategy and therefore, the risk-appetite of decision makers varies according to the organizational policy. Decision makers are broadly classified into three categories: Risk-averse decision maker, Risk-neutral decision maker and Risk-taking decision maker.

In classical economic theories, utility functions are used to model the decision maker's consideration for risks. According to this, a risk-neutral decision maker is the one whose aim is to maximize the expected benefits of investments by comparing the investment cost and the potential loss caused by possible security breaches [16]. A decision maker is considered risk-averse if the decision maker is reluctant to accept a bargain with an uncertain pay-off instead of a bargain with a more certain but lower expected payoff [16]. A decision maker having a preference for risk will be termed as risk-seeker. A risk-seeking decision maker has a preference for the uncertain path even though the uncertain and the certain path have the same expected value [16].

**(3) Financial Aspects:** From the upper management's perspective, a decision to invest in information security is driven by financial measures.

**(4) Non-Financial Aspects:** Besides financial aspects, there are a number of non-financial factors which drive the investments in information security. Increase in trustworthiness of an organization and meeting the compliance and regulatory guidelines are important non-financial factors encouraging investments in information security.

**(5) One-time Benefit:** Some of the investment models address only one-time benefit of an investment in information security. In this case the benefit of investment is realized immediately with the investment.

**(6) Recurring Benefit:** If an investment in an information security defence results in realization of benefits over a period of time then it is termed as recurring benefit. Some of the investment models for information security address only one-time benefits of investments while some are to address the recurring-benefits or both.

### 4.2. Content

The 'Content' is about the type of information which is required as an input to the 'Process'. The total cost and effectiveness of the information security controls is estimated and used as an input to the process phase.

Following parameters are derived from the 'Content' perspective to evaluate the information security investments:

**(1) Budget:** Usually the budget for information technology and security investments is limited. Thus, the models of information security investment are compared on the criteria of budget type, i.e. limited or unlimited budget.

**(2) Attack Type:** Some of the information security investment models are meant to address the single attack scenario while some are for multiple attacks. Therefore, a criterion of attack types is included in the frame-work to compare the models on the basis of single or multiple attack type.

**(3) Network Effects:** Information security investments generate positive and negative externalities. Therefore, this is an important aspect of information security which needs to be considered while deciding on investments in information security.

**(4) Input Type:** Some of the information security investment models rely on qualitative input while some on quantitative input. Therefore, we compare the models on the basis of input types: Qualitative or Quantitative.

**(5) Output Type:** The results of information security investment models vary in their nature. Some models present quantitative results while some present qualitative results.

**(6) Costs:** The costs associated with the information security controls can be classified as One-time cost, Recurring cost, direct cost, Indirect cost, Fixed cost, Variable cost, Sunk cost, and Recoverable cost. An example of a one-time cost is a purchase of a hardware product such as router. Recurring cost can be illustrated by a regular payment of license fee for a product. Acquisition, deployment and maintenance cost of products is an example of direct cost. Time and resources spent on some mandatory security or policy requirement such as changing a password at regular interval is an example of indirect cost. A pre-negotiated price or fee for a product or a service is an example of fixed cost. Variation in cost on distribution of products like security tokens to the customers is an example of variable costs. Costs incurred on training of staff to use a system (securely) or to comply with security policy of the company is a form of sunk cost for the organisation. Security tokens distributed to the customers are also a part of sunk cost. It is possible to recover some cost and an example of the recoverable cost is a repurposed use of routers or selling of firewall devices at a discounted price.

**(7) Depreciation of security:** Security attackers are regularly updating the software exploits and these software vulnerabilities are often shared on internet to equip the attackers with stronger attack capabilities. Thus, it is important to take information security investment decisions on the basis of depreciation of information security controls.

### 4.3. Process

The 'Process' presents the activities of planning the investments in information security defence mechanisms. The investments in information security can be based on proactive or reactive strategy. In the proposed framework 'Process' reflects the methodology of planning and evaluation.

Following parameters are derived from the 'Process' perspective to compare the models of investment in information:

**(1) Investment Approach:** From the commercial aspect of the business strategy, an organization may like to invest in information security controls in several stages. This strategy of investing in several phases allows an organization to utilize the available capital in the mean time for other revenue generating projects. The technical department of the organization may prefer to divide the investments in several phases to incorporate the latest technology/developments in the security systems. Thus, it becomes essential to compare the information security investment models on the basis of the type of investment strategy supported by the investment model. The investment models are either one-time investment model or a split-investment model.

**(2) Investment Strategy:** The strategy of investment in information security controls can broadly be divided into two categories, i.e. investments in information security controls before an attack has actually happened and then investments in information security controls after the organization has been attacked. These investment strategies can be termed as Proactive and Reactive strategy, respectively and therefore the information security investment models are compared on the basis of type of investment strategy supported by the models.

**(3) Investment Levels:** Information security investment models vary in their target application area. Some of the information security investment models are meant to address investments in a particular product or service specific security control while other investment models are to address the investments in organizational level security defence mechanisms. Therefore, the information security investment models are compared on the basis of their application level.

## 5. COMPARISON AND ANALYSIS OF INFORMATION SECURITY INVESTMENT MODELS

In this section, on the basis of the proposed framework an analysis and comparison between six information security investment models is presented.

Our classification of the models analysed is based on the claims made by the model's authors. That is, we make no judgments as to the extent of the correctness or applicability of the models.

### 5.1. Model-1: Gordon and Loeb Model

Gordon and Loeb [29] proposed an information security investment model to determine the optimal investment level for the protection of a single asset. To simplify the process, the model assumes that the decision maker is a risk-neutral decision maker. The model is a one-period model, where all the decisions and outcomes happen instantaneously. This means that the investments in information security controls are not divided into several phases. The model focuses only on the financial benefits of investments in the security and the non-financial benefits are ignored. The one-time costs and benefits of investments in the security system are taken into consideration; however the running costs and benefits of the investments are ignored. Further, the information on attacks and network effects are ignored in the model. The model does not specify the type of budget, i.e. whether the budget is limited or unlimited for information security investments; therefore, we believe that the model is suited for unlimited budget type.

The assumptions and description of the model are silent on conflict of interest between various stakeholders and there is no mention of cost associated with the conflict of interest between the stakeholders. Since the one-time cost is taken into account by the model, we can say that a partial direct cost of information security measures is also considered in the model. Though, the direct cost associated with the acquisition and deployment of security measures is considered but the recurring cost of maintenance is ignored. The description of the model does not mention anything about the indirect cost, so we assume that the indirect costs are not to be included in the model. As the model is supporting only one-time investment in information security, we further assume that the cost incurred on information security controls is fixed. The description of model is silent on the type of target security mechanisms; we are unable to decide on sunk cost, recoverable cost and depreciation of security.

### 5.2. Model-2: Mizzi Model

Mizzi [30] proposed a one-time investment model for investments in information security controls. Only the financial aspects of information security investments are taken into account and the non-financial aspects are left-out. The one-time cost as well as running-cost of information security controls is taken into account. The assumptions and description of the model do not con- vey any information about the type of decision maker. The description of the model does not mention anything about the type of budget, so we assume that the model is suited for unlimited budget type. The cost of attacker's effort is accounted in this model. Therefore, to increase the cost of attack for an attacker the cost incurred on information security controls is recurring in nature. On the other hand, a failure to upgrade the security with respect to change in attack scenario will count as depreciation in security (controls).

Various type of costs are included in the model, such as cost of fixing a vulnerability (which is a form of direct and fixed cost), annual maintenance costs (a type of recurring and variable cost). The cost of unavailability of an asset or system is accounted in the model and it is an example of indirect cost. The sunk cost is reflected in the form of man-hour costs considered by the model. The network effects are ignored in the Mizzi model and the model supports only single-asset investment. As the model considers the rebuilding cost of a compromised asset, the model supports the reactive investment strategy as well. The cost of conflict of interest is not included in the model.

### 5.3. Model-3: Al-Humaingani and Dunn Model

Al-Humaigani and Dunn [31] proposed a one-time investment model for investments in information security measures. The financial aspects are taken into account and the non-financial aspects are left out. The recurring costs and network effects are also ignored in the model. The cost of conflict of interest between the stakeholders is not considered in the model. The assumptions and description of the model have not specified the type of decision maker sup- ported by the model. The description of the model does not mention anything about the budget type; so we assume that the model is applicable in unlimited budget scenario. The model also ignores the information on attacks. As the model is based on the theory of return on security investments, the model supports a proactive strategy of investments in information security controls.

A range of costs are considered by the model, such as cost of procurement of security controls (which is form of direct and fixed cost), cost associated with additional hardware and facilities (which is a type of one-time investment), cost of training (which reflects the sunk and recurring cost), losses due to limitations placed on business (which is a form of indirect cost), cost of adopting a secured-by-design strategy (if the strategy is not updated regularly, this is partially a depreciation of security cost). Also, the need of regularly updating the security-by-design strategy is a kind of recurring cost.

An assessment of model-1, 2 and 3 against the proposed framework is presented in Table-1. The criteria which the model fulfils are marked with a '√', criteria which the model does not considers are marked as 'X' and a '-' in the table denotes absence of information on the criteria.

International Journal of Security, Privacy and Trust Management (IJSPTM) Vol 4, No 3/4, November 2015

| | | Evaluation Criteria | Gordon-Loeb Model | Mizzi Model | Al-Humaigani and Dunn |
|---|---|---|---|---|---|
| **Context** | Stakeholders | Cost of Conflict of Interest | X | X | X |
| | Decision Maker Type | Risk-Averse | X | - | - |
| | | Risk-Neutral | ✓ | - | - |
| | | Risk-Taker | X | - | - |
| | Benefits | Financial | ✓ | ✓ | ✓ |
| | | Non-Financial | X | X | X |
| | | One-time Benefit | ✓ | ✓ | ✓ |
| | | Recurring Benefit | X | X | X |
| **Content** | Budget Type | Limited Budget | X | X | X |
| | | Un-limited Budget | ✓ | ✓ | ✓ |
| | Attack Scenario | Single-Attack | X | X | X |
| | | Multiple-Attack | X | X | X |
| | Input Type | Qualitative Input | X | X | X |
| | | Quantitative Input | ✓ | ✓ | ✓ |
| | Output Type | Qualitative Output | X | X | X |
| | | Quantitative Output | ✓ | ✓ | ✓ |
| | Cost | One-time Cost | ✓ | ✓ | ✓ |
| | | Recurring Cost | X | ✓ | X |
| | | Direct Cost | ✓ | ✓ | ✓ |
| | | Indirect Cost | X | ✓ | ✓ |
| | | Fixed Cost | ✓ | ✓ | ✓ |
| | | Variable Cost | - | ✓ | ✓ |
| | | Sunk Cost | - | ✓ | ✓ |
| | | Recoverable Cost | - | - | ✓ |
| | | Depreciation of Security | - | ✓ | ✓ |
| | Network Effect | Externalities | X | X | X |
| **Process** | Investment Approach | One-time Investment | ✓ | ✓ | ✓ |
| | | Split Investment | X | X | X |
| | Investment Strategy | Proactive | ✓ | ✓ | ✓ |
| | | Reactive | X | ✓ | - |
| | Investment Level | Application/Service Specific | ✓ | ✓ | ✓ |
| | | Enterprise wide | - | - | X |

(Row label spanning left side: *Investing in Information Security from Strategic Perspective*)

Table 1. Assessment of Information Security Investment Models against the Proposed Framework

### 5.4. Model-4: Sonnenreich et al.

Sonnenreich et al. [32] proposed an information security investment model, which is similar to the traditional accounting approach of return on investment (ROI) and therefore they termed their model as return on security investment (ROSI). In ROSI model, investments are made as a whole, meaning it can- not be divided into several investment phases. The model considers only the financial measures and non-financial measures are ignored. The costs used in calculation of ROSI are not divided into different types. The benefits and costs incur only once. The information on attacks is not considered in the model and network effects are also ignored. There is no mention of budget type in the model and thus we believe that the model is applicable in unlimited budget scenario. The description of the model does not assume anything about the type of decision maker.

### 5.5. Model-5: Huang et al.

The security investment model proposed by Huang et al. [14] targets a risk-averse decision maker. The model is based on the assumption that in- vestments are made as a whole and cannot be split into several parts. The financial aspects are considered in the model; however the non-financial aspects are ignored. The model considers the one-time cost and benefits and recurring costs and benefits are not taken into consideration. The model is not dependent upon the information on attacks and network effects are also ignored. The budget type is not mentioned in the model description, thus we believe that the model targets the unlimited budget scenario.

International Journal of Security, Privacy and Trust Management (IJSPTM) Vol 4, No 3/4, November 2015

## 5.6. Cremonini and Martini

The information security investment model proposed by Cremonini and Martini [33] is similar to the model [32]. In addition to the return on investment approach, return on attacks was included in the Cremoini and Martini model. Return on attacks, allows evaluating the alternative defence mechanism from the attacker's perspective and the alternative with the highest disadvantage to the attacker is considered. Thus, there will always be a requirement to upgrade the defence mechanism as and when new vulnerabilities and exploits are discovered. A failure to timely upgrade the defence will lead to lower cost of attack for the attacker resulting into depreciation of security (control).

This model assumes that the investments are made as a whole and the investments cannot be split into several phases. The financial aspects are considered in the model and the non-financial aspects are left out. The costs and benefits incur only once. Contrary to the model [32], model [33] relies heavily on the information on attacks. However, the network effects are ignored. As the model does not specify the type of budget, we believe that it is applicable in unlimited budget scenario. Further, there is no mention of the type of decision maker.

An assessment of model-4, 5 and 6 against the proposed framework is presented in Table-2. The criteria which the model fulfils are marked with a '√', criteria which the model does not considers are marked as 'X' and a '-' in the table denotes absence of information on the criteria.

| | | | Evaluation Criteria | Sonnenreich et al. | Huang et al. | Cremonini and Martini |
|---|---|---|---|---|---|---|
| Investing in Information Security from Strategic Perspective | Context | Stakeholders | Cost of Conflict of Interest | X | X | X |
| | | Decision Maker Type | Risk-Averse | - | √ | - |
| | | | Risk-Neutral | - | X | - |
| | | | Risk-Taker | - | X | - |
| | | Benefits | Financial | √ | √ | √ |
| | | | Non-Financial | X | X | X |
| | | | One-time Benefit | √ | √ | √ |
| | | | Recurring Benefit | X | X | X |
| | Content | Budget Type | Limited Budget | X | X | X |
| | | | Un-limited Budget | √ | √ | √ |
| | | Attack Scenario | Single-Attack | X | X | √ |
| | | | Multiple-Attack | X | X | √ |
| | | Input Type | Qualitative Input | X | X | X |
| | | | Quantitative Input | √ | √ | √ |
| | | Output Type | Qualitative Output | X | X | X |
| | | | Quantitative Output | √ | √ | √ |
| | | Cost | One-time Cost | √ | √ | √ |
| | | | Recurring Cost | X | X | X |
| | | | Direct Cost | √ | √ | √ |
| | | | Indirect Cost | √ | √ | - |
| | | | Fixed Cost | √ | √ | √ |
| | | | Variable Cost | X | √ | X |
| | | | Sunk Cost | √ | - | - |
| | | | Recoverable Cost | √ | - | - |
| | | | Depreciation of Security | X | | √ |
| | | Network Effect | Externalities | X | X | X |
| | Process | Investment Approach | One-time Investment | √ | √ | √ |
| | | | Split Investment | X | X | X |
| | | Investment Strategy | Proactive | √ | √ | √ |
| | | | Reactive | X | X | X |
| | | Investment Level | Application/Service Specific | √ | √ | √ |
| | | | Enterprise wide | - | - | - |

Table 2. Assessment of Information Security Investment Models against the Proposed Framework

35

## 6. DISCUSSION

The framework presented in the paper simplifies the information security investment decisions and the information security strategy with the overall organizational strategy. The proposed framework considers the information security investments from a broader perspective and thus is able to overcome the limitation of frameworks presented in [17,18] where the decision making parameters are limited in nature. Further, following the iterative process of inclusion of new parameters under the CCP approach, the framework in this paper provides several new parameters derived from the assessment of information security investment models, which were not included in the framework presented in [16].

The framework presented in [17] is based on the fixed parameters of policy and security configuration management tool to evaluate the information security investment models. Thus, the framework is not suitable for evaluation of information security investments from the perspective of an organization's strategy. As the framework proposed by us is based on the CCP approach, it overcomes the limitation of the framework presented in [17] and is thus helpful in aligning the information security investment decisions with the organizational strategy.

The framework presented in [18] is based on a set of broad characteristics, and it does not provide a list of generic evaluation criteria for the in- formation security investment decisions. Thus, the framework is not a handy approach for the decision makers. Decision makers may find it difficult to link the framework's evaluation criteria with the organizational strategy. The framework proposed by us provides a list of parameters that are easy to link to the organizational strategy and thus the decision makers are in a better position to evaluate the investments in information security controls.

## 7. CONCLUSION

The framework presented in the paper simplifies the process of selection of information security investment model(s). The parameters for the comparison and analysis of information security investment models are derived from the 'Context', 'Content', and 'Process' perspective and are aimed at aligning the information security investments with the overall organizational strategy. The framework was applied to six information security investment models to analyse the applicability of these investment models with respect to the business strategy. This framework is useful for decision makers in identifying the most appropriate investment model or a combination of investment models. The framework is thus useful for assessment and prioritization of information security investments.

## 8. DIRECTIONS FOR FUTURE RESEARCH

The proposed framework is at an early stage of development. The framework needs to be applied to other information security investment models to assess the suitability of the models with respect the organizational strategy. Further, the sub-components of the three core components of the framework i.e. Context, Content, and Process, should be updated iteratively after the assessment of new information security investment models against the framework. Further, to test the workability of the framework in the real world scenario, the framework should be validated with the case studies and reviews of information security practitioners.

International Journal of Security, Privacy and Trust Management (IJSPTM) Vol 4, No 3/4, November 2015